\documentclass[12pt]{article}
\usepackage{amssymb}
\usepackage{amsfonts}
\usepackage{graphicx}
\usepackage{psfrag}
\usepackage{epsfig}
\newcommand{\sef}{\sin^2 \theta_{eff}^{lept}}

\newcommand{\sinteta}{\sin^2 \theta_{W}}

\newcommand{\ini}{\begin{equation}}
\newcommand{\fin}{\end{equation}}
\newcommand{\sms}{\hat{s}^2}
\newcommand{\cms}{\hat{c}^2}

\newcommand{\es}{s_{eff}^2}
\newcommand{\ec}{c_{eff}^2}
\newcommand{\drc}{\Delta\hat{r}}
\newcommand{\drhoc}{\Delta\hat{\rho}}
\newcommand{\drcw}{\Delta\hat{r}_W}
\newcommand{\dre}{\Delta r_{eff}}
\newcommand{\dr}{ \Delta r}
\newcommand{\bsli}{\begin{slide}}
\newcommand{\esli}{\end{slide}}
\newcommand{\bcen}{\begin{center}}
\newcommand{\ecen}{\end{center}}
\newcommand{\gl}{\Gamma_l}
\newcommand{\bite}{\begin{itemize}}
\newcommand{\eite}{\end{itemize}}
\newcommand{\bmath}{\begin{displaymath}}
\newcommand{\dah}{\Delta \alpha_h^{(5)}}
\newcommand{\asc}{\hat{\alpha}_s}
\newcommand{\emath}{\end{displaymath}}
\newcommand{\smallz}{{\scriptscriptstyle Z}} % small letters by paolo
\newcommand{\smallw}{{\scriptscriptstyle W}}

\newcommand{\mz}{M_\smallz}
\newcommand{\mw}{M_\smallw}
\newcommand{\msbar}{$\overline{MS}$}

\begin{document}

\hyphenation{re-nor-ma-li-za-tion}

\begin{flushright}
NYU-TH/02-09-01\\
hep-ph/0209079
\end{flushright}

%\begin{flushright}
%\end{flushright}
\vspace{0.5cm}
\begin{center}
{\Large \bf Thirty Years of Precision Electroweak Physics\ }\footnote{to appear in a future issue of 
Journal of Physics G.}\\
\vspace{0.2cm}
%{\Large\bf Eventual Comment.}\\
\vspace{0.9cm}
{\large Alberto Sirlin
%\footnote{e-mail: alberto.sirlin@nyu.edu}
}\\
\vspace{0.5cm}
{\it 
Department of Physics, 
New York University,\\
4 Washington Place, New York, NY 10003, USA.
}
\end{center}
\vspace{0.5cm}

\begin{center}
{\large         
\emph{J.J.~Sakurai Prize Talk\\
APS Meeting\\
Albuquerque, N.M., April 2002.}}\\

\vspace{2.0cm}

{\small
{\bf Abstract}

\begin{flushleft}
We discuss the development of the theory of electroweak radiative corrections and its role
in testing the Standard Model, predicting the top quark mass, constraining the Higgs boson mass,
and searching for deviations that may signal the presence of new physics.
\end{flushleft}
}

\vspace{1.0cm}

Dedicated to my unforgettable friends and collaborators: M.A.B.~B\'eg,
R.E.~Behrends, J.J.~Giambiagi, and J.J.~Sakurai, in memoriam.

\end{center}
\newpage

\section{Brief Historical Perspective} \label{section1}

The title of my talk, ``Thirty Years of Precision Electroweak Theory'',
refers to the fact that the Standard Model of Particle Physics (SM),
proposed originally by Weinberg, Salam, and Glashow \cite{a1}, emerged, with very important 
contribution by other physicists, in the period 1967-1974.

This theory has a basic property, shown by 't-Hooft, Veltman,
B.W.~Lee, Zinn-Justin, Becchi, Rouet and Stora,
and others, namely it is a renormalizable theory \cite{a2}.
This implies that it can be studied at the level of its quantum corrections 
by perturbative field theoretic methods, since the ultraviolet infinities 
encountered in the calculations can be absorbed as unobservable 
contributions to the masses and couplings of the theory.

At roughly the same time, a very ingenious and useful
method to regulate ultraviolet divergences, namely dimensional regularization,
was proposed almost simultaneously in three different parts of the world,
by 't-Hooft and Veltman in the Netherlands, by Bollini and Giambiagi in 
Argentina, and by Ashmore in Italy \cite{a3}. I first learned of 
dimensional regularization in a memorable conversation with Bollini and Giambiagi
that took place in Buenos Aires in January 1972.
Soon afterward, we found out that 't-Hooft and Veltman implemented this
method in the very important context of gauge theories. The
application of dimensional regularization to infrared divergences
was proposed soon afterward, by Gastmans and Meuldermans, and by
William (Bill) Marciano and me, while Bill was my research student at NYU \cite{a4}.
A bit later, Bill wrote a paper on his own extending 
the analysis to the regularization of mass singularities \cite{a5}. 
Dimensional regularization of infrared and mass singularities is widely employed 
at present, particularly in QCD calculations.

Once the renormalizability of the SM was recognized, it became natural
to explore this theory at the level of its quantum corrections. Already in the 
seventies there were a number of interesting developments:
\bite
\item[a)] the one-loop electroweak corrections (EWC) to $g-2$ date from that period.
\item[b)] Weinberg showed that there are no violations of ${\cal O}(\alpha)$ 
to parity and strangeness conservation in strong interaction amplitudes \cite{a6}.
\item[c)] Gaillard and Lee studied processes which are forbidden at the
tree level, but occur via loop effects, and showed that the GIM mechanism
generally suppresses neutral current amplitudes of ${\cal O}(G_F \alpha)$ \cite{a7}.
\item[d)] Veltman, and Chanowitz, Furman, and Hinchliffe discovered
that heavy particles do not generally decouple in the EWC of the SM,
and that a heavy top quark gives contributions of  ${\cal O}(G_F M_t^2)$ to
the $\rho$ parameter \cite{a8}.
\item[e)] Bollini, Giambiagi, and I studied the cancellation of
ultraviolet divergences in several fundamental natural relations
of the SM \cite{a9}.
\eite

My own main objective since the 70's has been the study of the EWC to allowed
processes, with the aim of bringing the theory into close contact with
precise experiments. The desiderata of these studies are:
\bite
\item[i)] To verify the SM at the level of its quantum corrections.
\item[ii)] To search for discrepancies or inferences that may signal 
the presence of new physics beyond the SM. 
\eite
These are essentially the objectives
of what is now called Precision Electroweak Physics.

At the time I felt that there was a problem that required urgent attention in 
order to test the tenability of the SM, namely the issue of Cabibbo 
universality or, in modern language, the test of the unitarity of the
CKM matrix. From studies in the framework of the Fermi theory
that preceded the SM, it was known that, in order to test Cabibbo 
universality, it is necessary to evaluate the radiative corrections
to muon decay and the Superallowed Fermi transitions in $\beta$-decay.
Nearly forty years ago it was shown that, to first order in $G_F$,
but all orders in $\alpha$, the photonic corrections to $\mu$-decay
are convergent in the Fermi V-A theory, after mass and charge renormalization \cite{a10}.
However, there was a big practical and conceptual problem: in the 
Fermi V-A theory the corrections to $\beta$-decay were known to be 
logarithmically divergent! 

Once the renormalizability of the SM was
recognized, it was apparent that the old conundrum
could be solved in the new framework. I argued with myself: if the theory
is renormalizable and I compute something physical, I should get a finite result!
Around 1974 I found
the answer in a simplified version of the SM, neglecting the strong
interactions \cite{a11}. However, a realistic evaluation of 
the EWC to $\beta$-decay is particularly
challenging, since one is dealing with a very low-energy-transfer
process affected by the strong interactions. 
Fortunately, and almost miraculously, their effect
can be controlled to a large extent using current algebra techniques
and associated Ward identities. The final result \cite{a12} was 
simple and encouraging: i) aside from some small, short-distance
QCD corrections, the result coincided with the regularized answer 
in the Fermi V-A theory, with the cutoff replaced by $M_Z$! ii) The corrections
turned out to be sizable. They are dominated by a large logarithmic term 
\bmath
3(\alpha/\pi)\ln{(M_Z/2 E_m)} \approx 3.4 \% \, ,
\emath
where $E_m={\cal O}(MeV)$ is the end-point of the positron spectrum in 
$\beta$-decay. Furthermore, such large corrections are phenomenologically
necessary to ensure, to good approximation, the tenability of
the SM in the analysis of universality \cite{a13}.
For me, this was the smoking gun of the SM at the level of its quantum corrections!

Towards the end of the 70's Bill and I thought that experimentalists would
probably search for the $W$ and $Z$ bosons and hopefully measure their masses.
It seemed a good idea to study at the loop level the relationship between
$M_W$, $M_Z$ and $G_F$, $\alpha$, as well as the other parameters of the SM,
such as $M_H$, $M_f$.
\begin{center}
How to do it?
\end{center} 
At the time we had precise knowledge of $G_F$ (which in my papers I had defined via
the muon lifetime evaluated in the Fermi V-A theory) and $\alpha$,
 and a less accurate knowledge of the electroweak mixing parameter $\sinteta$ from $\nu-N$
deep inelastic scattering via the neutral and charged currents. So it
became clear that it was necessary to evaluate the EWC to the last
two processes to establish the connection with  $\sinteta$, and to muon decay
to obtain the relation with $G_F$ and $\alpha$.

Since this required dealing with a number of processes involving neutral 
and charged currents, I strongly felt that the first step in the strategy 
should be to develop a simple method to renormalize the Electroweak Sector of 
the SM. I proposed this in a paper with a related title: ``Radiative Corrections
in the $SU(2)_L \times U(1)_Y$ theory: a Simple Renormalization Framework''
\cite{a14}.
This approach, with important subsequent contributions by other physicists,
is currently known as the on-shell scheme (OS). In the same paper, I 
applied the OS scheme to $\mu$ decay in the SM and introduced the EWC $\dr$,
whose significance I will briefly discuss later. In two subsequent papers with
Bill, the OS scheme was applied to study the EWC to $\nu-N$
deep inelastic scattering via the neutral and charged currents \cite{a15,a16}.
This trilogy of papers achieved our aim to establish contact with the expected 
measurements of $M_W$ and  $M_Z$ (which were carried out later). In fact,
the papers led to more accurate predictions of $M_W$, $M_Z$ using the OS relations \cite{a14,a17}
\ini \label{eq1}
s^2\,c^2 = \frac{A^2}{M_Z^2 (1 - \Delta r)}\quad  ,
\fin
\ini \label{eq2}
A^2 = \frac{\pi \alpha}{\sqrt{2} G_F} \, ,
\fin
\ini \label{eq3}
s^2 =\sin^2 \theta_W = 1-M_W^2/M_Z^2 \, ,
\fin
and the information from $\nu-N$ scattering.

During the 80's Bill and I also employed a hybrid \msbar\ scheme
where couplings are defined by \msbar\ subtractions 
but masses are still the physical ones.
It plays an important r\^ole in GUT predictions, which we also
studied \cite{a18}. The two schemes, OS and \msbar, were applied systematically
to additional important processes, such as $\nu$-lepton
scattering \cite{a19} and atomic parity violation \cite{a20}.

As experiments improved, the r\^ole of the EWC became more important
and Bill and I became part of a large collaboration, led
by Paul Langacker, whose aim was to elucidate the comparison between
theory and experiment. This culminated in a detailed review paper \cite{a21}. 
Some of the estimates of this analysis were
$M_W = 80.2 \pm 1.1\, \mbox{GeV}$, $M_Z = 91.6 \pm 0.9\, \mbox{GeV}$,
with central values within $0.2\, \mbox{GeV}$ and $0.4\, \mbox{GeV}$ from 
the current ones, respectively. We also obtained
$M_t < 180\, \mbox{GeV}\, @\, 90\% \, \mbox{CL}$  
for $M_H <  100\, \mbox{GeV}$. Over the years, Paul remained an
invaluable and very close collaborator.

Meanwhile, in the mid-eighties, a serious problem arose in the analysis of
the Superallowed Fermi transitions. Experiments on eight transitions
reached great accuracy and showed a significant departure from the
expectations of the conserved-vector-current hypothesis (CVC), which is an 
integral part of the SM. Simple theoretical arguments convinced me that the 
problem was related to the evaluation of the two-loop corrections
of  ${\cal O}(Z \alpha^2)$, which had been carried out numerically 
many years before.
My student Roberto Zucchini and I studied this correction analytically,
reviewed the analysis of the eight transitions in the light of our calculation, 
and found very good agreement with CVC \cite{a22}, a result that was confirmed by 
new numerical evaluations by Jaus and Rasche.

In the seventies and eighties I developed a fruitful collaboration with 
M.~A.~B.~B\'eg. Together, and often with other physicists, we wrote several 
papers and two extensive and, to some extent, pedagogical reviews on 
Gauge Theories of Weak Interactions \cite{b23a}. 

In the seventies I 
participated in an ambitious project, led by T.~D.~Lee, to study 
non-topological solitons in quantum field theories \cite{b23b}. Since my
post-doctoral years at Columbia, T.~D.~Lee, one of the great masters of
our discipline, has been for me a constant source of motivation and learning.

During the seventies, eighties, and nineties, I continued my close 
collaboration with Bill, with B.~A.~Kniehl, and several of my
students, former students, and post-doctoral associates: S.~Sarantakos,
S.~Bertolini, R.~Zucchini, G.~Degrassi, S.~Fanchiotti, P.~Gambino,
J.~Papavassiliou, K.~Philippides, M.~Passera, P.~A.~Grassi, A.~Ferroglia,
and  G.~Ossola.

Around 1989, LEP and SLC started operations, and FNL began the accurate measurement
of $M_W$. LEP soon determined $M_Z$ with great precision.
This prompted a change in strategy: $\alpha$, $G_F$, and $M_Z$ were
adopted as the basic input parameters, a great effort was made to study 
the observables at the $Z$ peak, namely the line shape and the various asymmetries
and widths measured at LEP and SLC, and there was a major improvement
in the comparison between theory and experiment.

\section{Input Parameters}

As I mentioned before, there are three very accurately determined 
quantities that play a special role as input parameters:
\bite
\item[i)] $\alpha = 1/137.03599959(38) (13)$,
$\delta \alpha = \pm 0.0037\, ppm$, 
 derived most precisely from $g_{(e)} -2$.
\item[ii)] $G_F \equiv G_{\mu} = (1.16637 \pm 0.00001) \times
10^{-5} \, GeV^{-2}$, 
$\delta G_F = \pm 9\, ppm$, where I defined $G_\mu$ from the muon lifetime 
using the finite photonic corrections of the V-A Fermi theory:
\begin{eqnarray} \label{eq4}
\delta &=& 1 + \frac{\alpha}{2 \pi}\,\left(\frac{25}{4} -\pi^2 \right)
\,\left[ 1 + \frac{2 \alpha}{3 \pi}\,\ln\left( \frac{m_{\mu}}{m_e}\right)\right]
\nonumber \\
& &
+ 6.701\,\left( \frac{\alpha}{\pi}\right)^2 + \cdots 
\end{eqnarray} 
The ${\cal O}(\alpha)$ term has been known for a long time \cite{a23}, 
the logarithmic term of ${\cal O}(\alpha^2)$
was derived several years later \cite{a24}, while the last term was evaluated very
recently \cite{a25}. It very nearly cancels 
the logarithmic term of ${\cal O}(\alpha^2)$.
Including very small 
${\cal O}(\alpha m_e^2 / m_{\mu}^2)$ contributions one has:
\ini \label{eq5}
\delta = 1 - 4.1995 \times 10^{-3} +1.5 \times 10^{-6} + \cdots \, ,
\fin
where the second and third terms stand for the one and two-loop contributions.
This reveals two interesting points: i) when the corrections are expressed in terms
of $\alpha$, as in Eq.(\ref{eq4}), by a fortuitous cancellation the 
${\cal O}(\alpha^2)$ effects are very small, and the original 
 ${\cal O}(\alpha)$ calculation turns out to be very accurate.
ii) It took about four decades to evolve from the one-loop to the complete two-loop
calculation! This is a sobering indication of how difficult
it may be to achieve the new frontier of complete two-loop 
calculations in the SM!
\item[iii)] $M_Z = 91.1875 \pm 0.0021 \, \mbox{GeV}$, $\delta M_Z = \pm 23\, ppm$
\eite

\section{Basic Electroweak Corrections} \label{section3}

There are a number of basic electroweak corrections that play an
important role in the analysis of the SM.

The EWC $\dr$ that appears in Eq.(\ref{eq1}) depends on the various 
physical parameters of the SM such as 
$\alpha$, $M_W$, $M_Z$, $M_H$, $M_f$, etc\ldots, where $M_f$ stands
for a generic fermion mass.
From a theoretical point of view, the significance of 
Eqs.(\ref{eq1}-\ref{eq3}) is that they provide the relation
between the physical parameters of the Fermi theory (low-energy effective
theory), namely $G_F$ and  $\alpha$, with those of the SM (underlying theory),
namely $\alpha$, $M_W$, $M_Z$, $M_H$, $M_f$\ldots, at the level of the
quantum corrections. Eqs.(\ref{eq1}-\ref{eq3}) are currently used to calculate
$M_W = M_W(M_H)$ leading to very sharp constraints on $M_H$. 
As is clear from Eqs.(\ref{eq1}-\ref{eq3}), $\Delta r$
is a physical observable.

Two other important relations are \cite{a26,a27} 
\ini \label{eq6}
\sms\,\cms =  \frac{A^2}{M_Z^2 (1 - 
\Delta \hat{r})}\, ,
\fin
\ini
\sms = \frac{A^2}{M_W^2 (1 - 
\Delta \hat{r}_W)}\, ,
\fin
where $\sms \equiv \sin^2\hat{\theta}_W (M_Z)$ is the electroweak
mixing parameter defined by modified minimal subtraction,
and evaluated at the scale $\mu = M_Z$.
It is employed in the $\overline{MS}$
scheme and plays a crucial r\^ole in GUT studies. 
$\Delta \hat{r}$ and $\Delta \hat{r}_W$ are the relevant EWC.
I introduced Eq.(\ref{eq6}) and evaluated $\Delta \hat{r}$
while visiting CERN in August of 1989, at the time LEP was starting operations.
By the end of Aug., LEP had measured $M_Z$ within $160\, \mbox{MeV}$.
Using $\Delta \hat{r}$ and the new $M_Z$ measurement, $\sms$ 
could be determined with significantly greater precision.
Further, the 
improved determination of $\sms$ was consistent with supersymmetric
Grand Unification \cite{a26}!

The \msbar\ and OS versions of the electroweak mixing parameter, namely 
$\sms$ and $s^2$, are related by \cite{a28}
\begin{eqnarray} \label{eq8}
\sms &=& s^2\,\left(1 + \frac{c^2}{s^2}\,\Delta \hat{\rho} \right) \, , \\ 
\Delta \hat{\rho} &=& Re\left[ \frac{A_{WW}(M_W^2)}{M_W^2}
-\frac{A_{ZZ}(M_Z^2)}{M_Z^2 \hat{\rho}}\right]_{\overline{MS}} \, , 
\end{eqnarray}
where $A_{WW}(q^2)$ and $A_{ZZ}(q^2)$ are the $W-W$ and $Z-Z$ 
transverse self-energies, $\hat{\rho} = (1- \Delta\hat{\rho})^{-1}$,
and $\overline{MS}$ denotes the $\overline{MS}$ renormalization and the
choice $\mu = M_Z$.           

Another important version of the electroweak mixing parameter is 
$\es \equiv \sin^2 \theta_{eff}^{lept}$, used by the 
Electroweak Working Group (EWWG)
to analyze the data at the $Z$ resonance.

The relations between $\es$ and $\sms$ and $s^2$ are given by \cite{a29}
\ini \label{eq10}
\es = Re \hat{k}_l (M_Z^2) \sms = Re k_l (M_Z^2) s^2 \, ,
\fin
where $\hat{k}_l(q^2)$ and $k_l(q^2)$ are electroweak form factors.
In particular, because of a fortuitous cancellation of effects,
$\mbox{Re}\hat{k}_l(M_Z^2)$ is very close to unity and 
$\es - \sms \approx 10^{-4}$.

It is also very convenient to employ the expression \cite{a30,a31}
\begin{eqnarray} \label{eq11}
\es\ \ec &=& \frac{A^2}{M_Z^2 (1 - \Delta r_{eff})}\, ,  \\ \label{eq12}
\Delta r_{eff} &=&  \Delta \hat{r} + \frac{e^2}{\es}\,\Delta \hat{k}\,
\left(1-\frac{\es}{\ec}\right)\,( 1 + x_t) + \cdots\, ,  
\end{eqnarray}
where $x_t = 3 G_{\mu} M_t^2 / 8 \sqrt{2} \pi^2$ is the 
leading contribution to $\Delta \hat{\rho}$.

The neutral current vertex of the $Z$ boson into an $f-\overline{f}$ pair has the form
\ini
<f\,\overline{f}| J_{\mu}^Z | 0 > = V_f(q^2)\,\overline{u}_f\,\gamma_{\mu}
\left[\frac{I_{3f} (1- \gamma_5)}{2} - \hat{k}_f(q^2)\sms Q_f \right] v_f \, ,
\fin
where $V_f(q^2)$, $\hat{k}_f(q^2)$, and its OS counterpart $k_f(q^2)$
are electroweak form factors, and $I_{3f}$ and $Q_f$ denote the third
component of weak isospin and the charge of fermion $f$.

The incorporation of QCD effects of ${\cal O}(\alpha \asc, \alpha \asc^2)$
in the basic EWC was studied with B.A.~Kniehl and my student Sergio
Fanchiotti \cite{a32}.

\section{Asymptotic Behaviors}

The basic corrections $\dr$, $\drc$, $\drcw$, $\dre$, $\drhoc$, $\hat{k}_f$,
\ldots have been studied in great detail by several groups. Here I can only point
out the asymptotic behaviors for large $M_t$, $M_H$ at the one loop level:
\begin{eqnarray} \label{eq14}
\Delta r & \sim & - \frac{3 \alpha}{16 \pi s^4}\, \frac{M_t^2}{M_Z^2}+
\frac{11 \alpha}{24 \pi s^2}\,\ln\left(\frac{M_H}{M_Z} \right)
+ \cdots \, ,\\ \label{eq15}
\Delta r_{eff} \approx \Delta \hat{r} & \sim & 
- \frac{3 \alpha}{16 \pi \sms \cms} \frac{M_t^2}{M_Z^2} +
\frac{ \alpha}{2 \pi \sms \cms} \left(\frac{5}{4} -\frac{3}{4}\cms  
\right)\ln\left(\frac{M_H}{M_Z} \right) + \cdots . \quad
\end{eqnarray} 
Eqs.(\ref{eq14}, \ref{eq15}) reveal a quadratic dependence on $M_t$,
a logarithmic dependence on $M_H$. Also, the asymptotic behaviors in $M_t$ 
and $M_H$ have opposite signs, which explains the well-known $M_t-M_H$
correlation.

The asymptotic behavior of the neutral current amplitude is
\ini \label{eq16}
\mbox{NC ampl.} \sim G_F/(1- x_t) \, , 
\fin
where $x_t$ is defined after Eq.~(\ref{eq12}).

Additional contributions to $\dr$ and $\dre$ lead to variations
\begin{eqnarray}
\delta M_W / M_W &\approx& - 0.205\, \delta(\Delta r) \, , \\
\delta \es / \es &\approx& 1.52\, \delta(\Delta r_{eff}) \, .
\end{eqnarray}

\section{The $M_t$ Prediction}

A very good example of the successful interplay between theory
and experiment was provided by the $M_t$ prediction and its subsequent 
measurement. Before 1995, the top quark could not be produced 
directly, but it was possible to estimate its mass because of its
virtual contributions to the EWC.
In Nov.~94, a global analysis by the EWWG led to the indirect determination
\ini
M_t = 178 \pm 11 ^{+18}_{-19}\, \mbox{GeV}\ , 
\fin 
where the central value corresponds to $M_H = 300 \, \mbox{GeV}$, 
the first error is experimental, and 
the second reflects the shift in the central value to 
$M_H = 65\, \mbox{GeV}\, (-19\, \mbox{GeV})$ or
$M_H = 1\, \mbox{TeV}\, (+18\, \mbox{GeV})$. This may be
compared with the current measurement
$(M_t)_{exp} = ( 174.3 \pm 5.1)\, \mbox{GeV}$.

This quite successful prediction was due to the quadratic $M_t$-dependence 
of the basic corrections, as illustrated in Eqs.(\ref{eq14},
\ref{eq15}, \ref{eq16}).

\section{Renormalization Schemes}

As discussed in Section~\ref{section1}, the EWC have been carried out in a number
of renormalization frameworks.
Two of the most frequently employed are: \\
%\bite
%\item[] 
On-Shell (OS) Scheme. It is ``very physical'', since it 
identifies renormalized couplings and 
masses with physical, scale-independent observables, such as 
$G_F$, $\alpha$, $\mz$, $\mw$, $M_H$, $M_f$, \ldots \\
%\item[] 
\msbar\ Scheme. It has very good convergence properties.
This is related to the fact that in this scheme one essentially 
subtracts the pole terms and, therefore, the
calculations follow closely the structure of the unrenormalized theory. 
As a consequence, it avoids large finite corrections frequently 
induced by renormalization. 
It employs inherently scale-dependent couplings such as $\alpha (\mu)$,
$\sms (\mu)$, which play a crucial r\^ole in the analysis of Grand Unification.
On the other hand, this leads to a residual scale dependence 
in the calculation of observables in finite orders of
perturbation theory. The choice $\mu = M_Z$ is frequently made. 
%\item[]

Very recently, a novel approach was proposed with my students Ferroglia
 and Ossola \cite{a31,a33}: \\
 Effective Scheme. It
shares the good convergence properties of the \msbar\ approach, but 
the calculation of observables in this scheme is strictly 
scale independent in finite orders. It employs scale-independent
quantities such as $\es$, $G_F$, $M_Z^2$ as basic parameters.
The reason that the Effective Scheme shares the good convergence properties
of the \msbar\ approach is related to the fact that, as mentioned before,
$\sef$ and $\sms(M_Z)$ are numerically very close (Cf. discussion
after Eq.(\ref{eq10})).
%\eite

\section{The running of $\alpha$}
A very important contribution to the EWC is due to the running of
$\alpha$ to the $\mz$ scale (vacuum polarization contributions):
\ini
\alpha (M_Z)/\alpha = 1/(1-\Delta \alpha) \, .
\fin
The light quarks' contribution ($u$-$b$) is evaluated using 
dispersion relations
and the experimental cross section for
$e^+ + e^- \to \mbox{hadrons}$ at low
$\sqrt{s}$, and perturbative QCD (PQCD) at large $\sqrt{s}$.
Recently, ``theory driven'' calculations claim to reduce the 
error by using PQCD down to low $\sqrt{s}$ values.

In the Winter 2002 analysis \cite{a34}, the EWWG employs two determinations:
\ini \label{eq21a}
\dah = 0.02761 \pm 0.00036 \, ,
\fin
and
\ini \label{eq21b}
\dah = 0.02747 \pm 0.00012 \, .
\fin
The leptonic contribution is
\ini \label{eq21c}
\Delta \alpha_l = 0.03150 \, .
\fin

\section{Evidence for Electroweak Corrections}

\bite
\item[A)] Evidence for EWC beyond the running of $\alpha$ \cite{a35}.
It can be obtained by measuring $\Delta r$. 
Using $(M_W)_{exp} = 80.451 \pm 0.033 \, \mbox{GeV}$ \cite{a34}, and 
Eqs.(\ref{eq1}-\ref{eq3}) one finds
$(\Delta r)_{exp} = 0.03107 \pm 0.00200$.
The contribution to $\Delta r$ from the running of $\alpha$ is
$\Delta \alpha = 0.05911 \pm 0.00036$, where I used 
Eqs.(\ref{eq21a},\ref{eq21c}). The
EWC not associated with $\Delta \alpha$ is  
$(\Delta r)_{exp} - \Delta \alpha = - 0.02804 \pm 0.00203$,
which differs from $0$ by $13.8\ \sigma$!
A similar result is obtained by comparing $(\es)_{exp} = 0.23149 \pm 0.00017$ \cite{a34}
and $(s^2)_{exp} = 0.22162 \pm 0.00064$. The difference is $0.00987 \pm 0.00066$
or $14.9\ \sigma$! And it is due to EWC not involving $\Delta \alpha$. In 
fact, this difference is dominated by the correction $c^2 \drhoc$ in 
Eq.(\ref{eq8}). 
\item[B)] Evidence for Electroweak Bosonic Correction (EWBC) \cite{a36}.
They include loops involving the bosonic sector, $W$'s, $Z$, H.
They are subleading numerically, but very important conceptually.
Evidence for these correction can be found by measuring 
$(\Delta r)_{eff}$. Using $(\es)_{exp} = 0.23149 \pm 0.00017$
and Eq.(\ref{eq11}),
we find $(\Delta r_{eff})_{exp} = 0.06047 \pm 0.00048$. 
Subtracting the contribution of the EWBC, but retaining the fermionic 
corrections, the theoretical value is
 $(\Delta r_{eff})_{theor}^{subtr} = 0.05106 \pm 0.00083$.
The difference is $(\Delta r_{eff})_{exp} - (\Delta r_{eff})_{theor}^{subtr} = 
0.00941 \pm 0.00096$, a $9.8\  \sigma$ effect!
\eite

\section{Theoretical Pursuit of the Higgs Boson}

The Higgs boson is the
fundamental missing piece of the SM!
With $M_t$ measured, to what extent can $M_H$ be constrained?
For large $M_H$, the EWC contain contributions proportional to $\ln{M_H/M_Z}$. 
We need precise calculations! 
Theorists distinguish two classes of errors:
\bite
\item[1)] parametric, such as $\delta M_t$, $\delta \dah$, $\delta \es$, \ldots
\item[2)] uncertainties due to the truncation of the perturbative series
(i.e. uncalculated higher order effects). What is the status of the higher order 
corrections?
Contributions of ${\cal O}(\alpha)$, ${\cal O}\left(\alpha \log{M_Z/M_f} \right)^n$,
and \, ${\cal O}\left(\alpha^2 \log{M_Z/M_f} \right)$ were analyzed during the 
period 1979-84. Those of ${\cal O}\left(\alpha^2 (M_t/M_W)^4 \right)$,
${\cal O}( \alpha \alpha_s)$, and 
${\cal O}\left( \alpha \alpha_s^2 (M_t/M_W)^2 \right)$ were studied from
the late 80's to the middle 90's.
\eite

Of more recent vintage are the corrections of 
${\cal O}\left(\alpha^2 (M_t/M_W)^2 \right)$.
Large $M_t$ expansions were employed to evaluate the irreducible
contributions of this order to the basic corrections,
which were then incorporated in the calculation of $\es$
and $M_W$, as functions of $M_H$, in three schemes:
\msbar, and two versions, OSI and OSII, of the OS scheme,
with two different implementations of the QCD corrections
\cite{a37}.
A large reduction was found in the scheme and residual scale
dependences. Maximal variations, among the schemes, 
for given $M_H$, amounted to 
$\Delta \es \approx 3 \times 10^{-5}$ and $\Delta M_W \approx 2\ \mbox{MeV}$.
Including additional QCD uncertainties: 
$\Delta \es \approx 6 \times 10^{-5}$ and $\Delta M_W \approx 7\ \mbox{MeV}$.
In the case of $M_W$, the results can be compared with
important new calculations that include all two-loop
contributions to $\Delta r$ that contain a fermion loop
\cite{a38}. Again one finds
$\Delta M_W \approx 7\ \mbox{MeV}$.
The study of the ${\cal O}\left( \alpha^2 M_t^2/M_W^2 \right)$
contributions has been extended to the partial widths
$\Gamma_f (f \neq b)$ of the $Z$ \cite{a39}
and to the Effective Scheme \cite{a31,a33,a40}.
The incorporation of the  ${\cal O}\left( \alpha^2 M_t^2/M_W^2 \right)$
had also a felicitous consequence: the $95\% $ CL upper bound $M_H^{(95)}$ 
was reduced by $\approx 30 \% $ \cite{a41}.

\section{Simple Formulae for $\es$, $M_W$, $\Gamma_l$}

Very simple formulae, that reproduce accurately
the numerical results of the detailed codes in the range
$20\, \mbox{GeV} \le M_H \le 300\, \mbox{GeV}$, have been recently
presented \cite{a40}. They are of the form
\begin{eqnarray} \label{eq22a}
\es = & (\es)_0 + c_1 A_1 + c_5 A_1^2 + c_2 A_2 - c_3 A_3 + c_4 A_4, \\
\label{eq22b}
M_W = & M_W^0 - d_1 A_1 - d_5 A_1^2 - d_2 A_2 + d_3 A_3 - d_4 A_4, \\
\label{eq22c}
\Gamma_l = & \Gamma_l^0 - g_1\,A_1 - g_5\,A_1^2  - g_2\,A_2 + g_3\,A_3 - 
g_4\,A_4 \, , 
\end{eqnarray}
where $\Gamma_l$ is the leptonic
partial width of the $Z$, $c_i$, $d_i$, $g_i$
 $(i=1-5)$ are constants
 given in Ref.~\cite{a40}, and
\begin{eqnarray}
A_1 \equiv \ln \left( M_H / 100 \, \mbox{GeV}\right) \, ,\quad &
A_2 \equiv \left[ \Delta \alpha_h^{(5)}/0.02761 \right]  - 1 \, , \nonumber \\
A_3 \equiv \left( M_t / 174.3 \, \mbox{GeV}\right)^2 -1 \, , \quad &
A_4 \equiv \left[ \alpha_s(M_Z) / 0.118 \right] -1 \, . \label{eq22d}
\end{eqnarray} 
In constructing these expressions, the input values 
$M_t = 174.3 \pm 5.1 \, \mbox{GeV}$, 
$\Delta \alpha_h^{(5)} = 0.02761 \pm 0.00036$, 
$\alpha_s \left( M_Z\right) = 0.118 \pm 0.002$,
were employed \cite{a34}.
A very useful feature is that Eqs.(\ref{eq22a}-\ref{eq22d}) retain
their accuracy over the rather large range 
$0.0272\, \le \dah \le 0.0283\ $ that encompasses the recent calculations.

We now discuss some instructive physical applications
of  Eqs.(\ref{eq22a}-\ref{eq22d}).
\bite
\item[i)] Using only Eq.(\ref{eq22a}) 
with $(\es)_{exp} = 0.23149 \pm 0.00017$ \cite{a34}, one finds
\bmath 
M_H = 124_{-52}^{+82}\, \mbox{GeV};\quad M_H^{95} = 280 \, \mbox{GeV} ,
\emath
where $M_H^{95}$ stands for the $95 \% \ \mbox{CL}$ upper bound.
\item[ii)]
Using only Eq.~(\ref{eq22b}) with $(M_W)_{exp} = 80.451 \pm 0.033 \, \mbox{GeV}$ \cite{a34},
one obtains
\bmath
M_H = 23_{-23}^{+49}\,\mbox{GeV};\quad M_H^{95} = 122 \, \mbox{GeV} .
\emath 
Thus, at present,
$(M_W)_{exp}$ constrains $M_H$ much more sharply than $\es$!
In fact, the above $M_H$ value is well below the direct 
exclusion bound $M_H > 114 \, \mbox{GeV}$ ($95\%$ CL).
\item[iii)] Use $A_1$, derived from $(\es)_{exp}$ and Eq.~(\ref{eq22a}),
to predict $M_W$ via Eq.(\ref{eq22b}). This leads to: 
\bmath
\left( M_W \right)_{indir.} = \left( 80.374 \pm 0.025 \right)\, \mbox{GeV} \, ,
\emath
which differs from $(M_W)_{exp}$ by $1.86\sigma$.
The above value is close to  
$(M_W)_{indir.} = 80.379 \pm 0.023\, \mbox{GeV}$, obtained
in the global analysis \cite{a34}.
\item[iv)] Using simultaneously Eqs.(\ref{eq22a}-\ref{eq22c}) with
$(\es)_{exp}$, $(M_W)_{exp}$, $(\gl)_{exp}$, one obtains
\bmath 
M_H = 97_{-41}^{+66}\, \mbox{GeV}\,;\quad  M_H^{95} = 223 \, \mbox{GeV}\, ,
\emath 
that may be compared with
\bmath
M_H = 85_{-34}^{+54}\, \mbox{GeV}\,;\quad  M_H^{95} = 196 \, \mbox{GeV}
\emath
in the recent EWWG fit \cite{a34}.
\eite

The current $(\es)_{exp}$ determination by the EWWG has  
$\chi^2/\mbox{{\small d.o.f.}}=10.6/5$, which corresponds to a  CL of only $6\%$.
There is an intriguing dichotomy: from the leptonic observables
($A_l(SLD)$, $A_l(P_\tau)$,
$A_{fb}^{(0,l)}$) one finds $(\es)_{l}=0.23113 \pm 0.00021$, while the 
hadronic measurements ($A_{fb}^{(0,b)}$, $A_{fb}^{(0,c)}$,
$<Q_{fb}>$) lead to  $(\es)_{h}=0.23220 \pm 0.00029$.
Thus, there is a $3\ \sigma$ difference between the determinations of $\es$
from the leptonic and hadronic sectors!
It is also interesting to note that
 $(\es)_{l}$  leads to 
\bmath
M_H = 59_{-29}^{+50}\, \mbox{GeV}\,;\quad  M_H^{95} = 158 \, \mbox{GeV}\, ,
\emath
which are closer to the $M_H$ values derived from $(M_W)_{exp}$.

If $(\es)_{h} - (\es)_{l}$ is due to a
statistical fluctuation, one possibility is to 
increase the error by $[\chi^2/\mbox{d.o.f.}]^{1/2}$, according to 
the PDG prescription. This results in $\es = 0.23149 \pm 0.00025$.
Interestingly, increasing the error in $\es$ leads to a smaller $M_H^{(95)}$
in the combined $\es$-$M_W$-$\Gamma_l$ analysis: 
($223\, \mbox{GeV} \to 201\, \mbox{GeV}$)!
The reason is that increasing the $\es$ error gives enhanced 
weight to the $M_W$ input, which prefers a smaller $M_H$ value.

If $(\es)_{h} - (\es)_{l}$ is due to new physics involving the 
($t$,$b$) generation, a substantial, tree-level change 
in the $Zb_R\overline{b}_R$ coupling would be required \cite{a42,a43}.
Very recently, it has been pointed out that if the $(\es)_{h}$ and $(\es)_{l}$
discrepancy were to settle on the leptonic value, a scenario with
light ${\tilde \nu}$, ${\tilde l}$, and  ${\tilde g}$ would improve 
the agreement with the 
electroweak data and the direct lower bound on $M_H$ \cite{a44}.
      
\section{Global Fit}

The SM describes rather well a large number of observables. Recent
fits lead to:
\bmath 
M_H = 85_{-34}^{+54}\, \mbox{GeV}\,;\quad  M_H^{95} = 196 \, \mbox{GeV} \quad \mbox{\cite{a34}} \, ,
\emath
\bmath 
M_H = 90_{-33}^{+50}\, \mbox{GeV}\,;\quad  M_H^{95} = 197 \, \mbox{GeV} \quad \mbox{\cite{a45}} \, .
\emath
There are no major deviations from the SM fit.
For instance, in the EWWG group
analysis, the largest differences are $3\ \sigma$ for  
$\sin^2 \theta_{W}(\nu N)$, $-2.64\ \sigma$ for $A_{fb}^{(0,b)}$,
$1.73\ \sigma$ for $M_W$, $1.63\ \sigma$ for $\sigma^0_{had}$, and 
$1.50\ \sigma$ for $A_l(SLD)$.

Thus, there are no major disagreements or compelling 
signals for new physics.

Nonetheless, it is very important to explore for new physics. For example,
if the central values of $M_t$ and $M_W$ remain as they are now, but the 
errors shrink sharply as expected at Tevatron/LHC or even much better 
at LC+GigaZ, a discrepancy would be established with the SM, that can 
be accommodated in the MSSM. It is also very important to remember 
that $M_H < 135 \, \mbox{GeV}$ in the MSSM.
As emphasized by Bill in his talk, the measurement and analysis of the muon 
anomaly $g_\mu - 2$ is of particular interest at present. If a conclusive deviation
from the SM prediction were established, an intriguing possibility would be the presence 
of supersymmetric contributions!

\section{Precision Studies, Quantum Field Theory and Fundamental 
Physical Concepts}

The foundations of the SM are firmly rooted in major developments 
in Quantum Field Theory: Yang-Mills Theories, their quantization and 
renormalizability; BRS symmetry; spontaneous symmetry breaking;
renormalization schemes and their implementation;
new techniques of computation, etc\ldots

Precision studies have also led to
unexpected byproducts. I mention two in which I was involved:
\bite
\item[i)] The discovery of the cancellation of mass-singularities 
in integrated transition probabilities (first paper in Ref.~\cite{a23}).
This was an important motivation for the Kinoshita-Lee-Nauenberg (KNL)
Theorem.
\item[ii)] The elucidation of the concepts of mass and width of unstable 
particles \cite{a46} and Partial Widths \cite{a47}. In 1991
I realized that, in the context of gauge theories,  the 
conventional on-shell definitions of mass and width
are gauge dependent in next-to-next-to-leading order, and proposed
to solve this severe conceptual and practical problem in terms
of definitions based on the complex-valued position of the propagator's pole.
\eite

\section*{Concluding Remarks}

With improving experimental precision, the study of electroweak
and QCD corrections plays an increasingly important role.

The modern era of these studies, in the framework of the Fermi theory,
started in the mid-fifties, in collaboration with R.~E.~Behrends
and R.~J.~Finkelstein, who was our mentor  \cite{a48}. 
Since that time until the emergence of the SM, the significance of the 
problem of universality attracted the attention of several first rate 
theorists.

However, at any given moment, the number of physicists engaged in these 
studies was very limited: you could count us with the fingers of one hand!

The emergence of the SM created a new theoretical framework where these
studies can be carried out in a theoretically 
consistent manner. At the same time
we got lucky: experimental physics moved in the direction of precision 
electroweak physics and a rich phenomenology emerged! In fact, it 
is very likely that precision electroweak physics 
will continue to be an important component
in the future development of our science.

For me, a particularly rewarding experience is to walk into a room
at a Conference or a Workshop and see dozens of talented young theorists
(some of them my own students and collaborators) working in this 
frontier area of Physics!

\section*{Acknowledgments}

The author is indebted to J.~Erler for illuminating communications.
This work was supported in part by NSF Grant No.~PHY-0070787.

\end{document}